\documentclass[%
reprint,
superscriptaddress,
amsmath,amssymb,
aps,
prb,
]{revtex4-2}

\usepackage{booktabs}
\usepackage{multirow}
\usepackage{color}
\usepackage{graphicx}
\usepackage{dcolumn}
\usepackage{bm}
\usepackage{hyperref}
\usepackage[mathlines]{lineno}
\usepackage{tabularx}

\begin{document}
	
	
	\title{Laser-engineered $\Gamma$-point Topology in Trigonal Bismuthene}
	
	\author{Zhe Li}%
	\affiliation{%
		Beijing National Laboratory for Condensed Matter Physics, and Institute of Physics, Chinese Academy of Sciences, Beijing 100190, China
	}%

	\author{Haijun Cao}
	\affiliation{%
		Beijing National Laboratory for Condensed Matter Physics, and Institute of Physics, Chinese Academy of Sciences, Beijing 100190, China
	}%
	
	\affiliation{%
		University of Chinese Academy of Sciences, Beijing 100049, China
	}
	
	\author{Sheng Meng}
	\affiliation{%
		Beijing National Laboratory for Condensed Matter Physics, and Institute of Physics, Chinese Academy of Sciences, Beijing 100190, China
	}%
	
	\affiliation{%
		University of Chinese Academy of Sciences, Beijing 100049, China
	}
	
	\affiliation{%
		Songshan Lake Materials Laboratory, Dongguan, Guangdong 523808, China\\
		{\normalfont Corresponding author: Sheng Meng.	}\textbf{\normalfont e-mail:} {\normalfont smeng@iphy.ac.cn}
	}
	
	\date{\today}

	\begin{abstract}
		The $\Gamma$-point topology represents a significant segment in the family of topological insulators. Here we provide a comprehensive prediction of light-induced $\Gamma$-point-based topological manipulation in trigonal bismuthene and its derivatives. Our findings unveil a two-stage process of topological phase transitions (TPT) as the laser intensity increases. Initially, a quantum-spin-Hall or metallic state transitions to a quantum-anomalous-Hall (QAH) state ($C$ = $\pm$3), followed by another TPT that yields a compensated Chern-insulating state ($C$ = 0). The trigonal warping model accounts for these states, describing the $C_{3z}$-rotational band-inversion process, which is determined by $\pm$1 orders of replica bands. Notably, this high Chern-number QAH state persists over a broad range of laser parameters, maintaining functionality beyond room temperature as evidenced by the large global gaps ($\geq$ 60 meV). Our work provides a comprehensive roadmap towards the designer $\Gamma$-point topology under laser excitation, facilitating applications of artificial topological materials.
	\end{abstract}
	
	\maketitle
	
	The $\Gamma$-point-based topology with $C_{3z}$-rotational symmetry plays a fundamental role in traditional building blocks, including trigonal bismuthene, TlBiSe$_2$, Bi(Sb)$_2$Te$_3$, Bi$_2$Se$_3$, and the MnBi$_2$Te$_4$-family of materials \cite{reis2017bismuthene,liu2011stable,pumera20172d,saini2022dft,xue2024tunable,zhang2009topological,liu2010oscillatory,liu2010model,chen2009experimental,niu2011quantum,singh2012topological,singh2025thermal,chang2013experimental,li2019intrinsic,otrokov2019unique,zhang2019topological,gong2019experimental,deng2020quantum,liu2020robust,ge2020high,li2020tunable,bai2024quantized,tang2023intrinsic,li2024realization,li2025universally,grutter2021magnetic,li2024multimechanism,hong2025designing}. The electronic structure of these materials can be nicely characterized by the prototypic four-band model proposed by Ru \textit{et al}. \cite{yu2010quantized}, which describes the critical microscopic mechanisms that govern topological phase transitions (TPTs). Specifically, the introduction of significant internal magnetism induces Zeeman splitting in the two spin components, while the emergence of spin-orbit coupling (SOC) subsequently opens the Chern insulating gap. As a representative of two-dimensional (2D) quantum spin Hall (QSH) building blocks with $\Gamma$-point-based topology, trigonal bismuthene, when placed on a magnetic substrate, exhibits a high-Chern-number quantum anomalous Hall (QAH) effect, with $C$ = $\pm$3 for out-of-plane magnetism and $C$ = $\pm$1 for in-plane magnetism along specific orientations \cite{xue2024tunable}. Contributed by the $p_{\pm}$ orbitals of Bi around the $\Gamma$ point in the Brillouin zone, the trigonal warping term results in high Chern numbers with significant tunability. This flexible tunability is also prophesied in materials such as PdSbO$_3$, NiAsO$_3$, and NiBiO$_3$ \cite{li2022chern,wu2023robust}. The theoretical and experimental investigations mentioned above address a substantial proportion of the challenges in this field.

Manipulating electronically excited states through Floquet engineering offers a continuously modulable technique for achieving a variety of novel topological properties \cite{liu2023floquet,zhan2024perspective,oka2009photovoltaic,bao2022light,liu2018photoinduced,hubener2017creating,zhan2023floquet,zhu2023floquet,qin2023light,wan2024photoinduced,liu2022high,li2025light,zhang2025quantum,wang2013observation,mahmood2016selective,choi2025observation,merboldt2025observation,zhou2023pseudospin,qiu2018ultrafast,bielinski2025floquet,mciver2020light,usaj2014irradiated,liu2023floquet-b,li2024floquet,bukov2015universal,blanes2009magnus}. By reshaping electronic bands through interactions with an infinite number of replica bands \cite{liu2023floquet}, Floquet engineering possesses a powerful capability that is not available in the ground-state regime. By utilizing circularly polarized light (CPL) or elliptically polarized light, time-reversal symmetry is broken, yielding multiple stages of TPTs. Numerous theoretical predictions support this capability, including the transitions between different types of semimetal phases in bulk black phosphorus \cite{liu2018photoinduced}, the modulation of valley-based Chern numbers in monolayer VSi$_2$N$_4$ \cite{zhan2023floquet}, the reversal of Chern number chirality in thin-film MnBi$_2$Te$_4$ \cite{zhu2023floquet}, the emergence of QAH states in thin-film Bi$_2$Te$_3$ \cite{qin2023light}, TPTs from high-order topology to QAH states \cite{wan2024photoinduced}, and Floquet-Chern flatbands induced in a Kagome material \cite{liu2023floquet,liu2022high}, among others \cite{li2025light,zhang2025quantum,wang2013observation,mahmood2016selective,choi2025observation,zhou2023pseudospin,qiu2018ultrafast,bielinski2025floquet,mciver2020light,usaj2014irradiated,liu2023floquet-b,li2024floquet}. Remarkably, Floquet engineering facilitates TPTs from QSH to QAH states in typical 2D building blocks, including group-IV Xenes \cite{li2025light} and square-lattice monolayer bismuth \cite{zhang2025quantum}. In the former case, large global gaps emerge without the limitations imposed by magnetism, enhancing the potential for achieving above-room-temperature QAH states across a wide range of laser parameters \cite{li2025light}. Employing time-resolved angle-resolved photoelectron spectroscopy, i.e. tr-ARPES, Floquet-Bloch bands with $\pm$1 orders of replica have been observed in Bi$_2$Se$_3$ \cite{wang2013observation,mahmood2016selective}, graphene \cite{choi2025observation,merboldt2025observation}, black phosphorus \cite{zhou2023pseudospin,qiu2018ultrafast} and MnBi$_2$Te$_4$ \cite{bielinski2025floquet}. Moreover, an anomalous Hall conductance of $\sigma_{xy}\ =\ (1.8\pm0.4) \frac{e^2}{h}$, which is close to the quantized value, has been observed in  graphene under laser illumination \cite{mciver2020light}, further highlighting the significance of Floquet engineering in the topological states of matter. Despite these accomplishments, the understanding and characterization of Floquet-Bloch states in most traditional $\Gamma$-point-based $C_{3z}$-rotational topology remains elusive. 
	
	In this work, we comprehensively investigate the $\Gamma$-point-based topological properties of trigonal bismuthene and its derivatives (bismuthene-antimonene and antimonene) under the incidence of CPL, focusing on the hybridization between the $\pm$1 orders of replica bands. Our investigation reveals a two-stage TPT process from a QSH state to a high Chern-number QAH state ($C$ = $\pm$3), followed by another TPT into a compensated Chern-insulating phase ($C$ = 0). The trigonal warping model \cite{fu2009hexagonal,liu2013plane,ren2016quantum,zhong2017plane} accounts for these light-induced topological states, characterized by nontrivial Berry curvature points that obey  $C_{3z}$-rotational symmetry. Furthermore, we find that the ground state of the QSH system is not a prerequisite for the emergence of electronically excited high Chern-number QAH states. For instance, laser pulses with a moderate photon energy can also induce QAH states ($C$ = $\pm$3) in topologically trivial bismuthene-antimonene. In addition, decreasing the incident angle introduces another $\Gamma$-point positioned QAH state ($C$ = $\pm$1) at specific ranges of laser intensities. For the first time, we present a comprehensive roadmap for engineering $\Gamma$-point-based $C_{3z}$-rotational topology rooted in the first order of Floquet-Bloch states, addressing this critical issue in traditional topological building-blocks and paving the way for future theoretical investigations and experimental applications of the resultant topological states.
		
		
	\vspace{1.5em} 
	\noindent
	\textbf{\large Results}
	
	\noindent
	
	\noindent
	Similar to group-IV Xenes, group-V bismuthene-family materials exhibit a low-buckled trigonal structure with the space group $P\bar{3}m1$ (No. 164). Nevertheless, due to differences in the occupancy of $p$-electrons and the absence of dangling bonds, bismuthene-family material elimiates the topological characteristics around the $K$ ($K’$) valleys and results in extraordinarily large band gaps. Considering the topologically nontrivial features around the zone center $\Gamma$-point in the reciprocal space of plumbene \cite{zhang2021selective} and stanene \cite{xu2013large} and the fairly strong SOC strengths, bismuthene also functions as a large-gap ($\sim$0.48 eV) 2D QSH insulator. Its topology arises from band inversion between the Bi 6$p_z$ and 6$p_{\pm}$ states around the $\Gamma$-point (as depicted in Supplementary Fig. 1 of Supplementary Note 1). According to the insights of the Floquet-Bloch theorem, the incidence of light generates infinite order of replica bands, leading to band reformation as a result of band hybridization between these Floquet-Bloch states. Here, we define the light intensity as a reciprocal vector $A = \frac{eA_{0}}{\hbar c}$, which is in unit of Å$^{-1}$. It is challenging to directly assess potential TPTs determined by zero-order bands, as ultrahigh light intensities ($\geq$ 0.2 Å$^{-1}$, see Supplementary Fig. 2) are required in the high-frequency regime, with details shown in Supplementary Note 2. 
    In this regime of high laser intensity, thermal effects become non-negligible, and structural damage may occur. Therefore, we include the $\pm$1 orders of replica bands, which are post-processed after hybridization with the zero-order and $\pm$2 orders of replica bands.

\begin{figure*}
	\centering
	\includegraphics[width=0.8\linewidth]{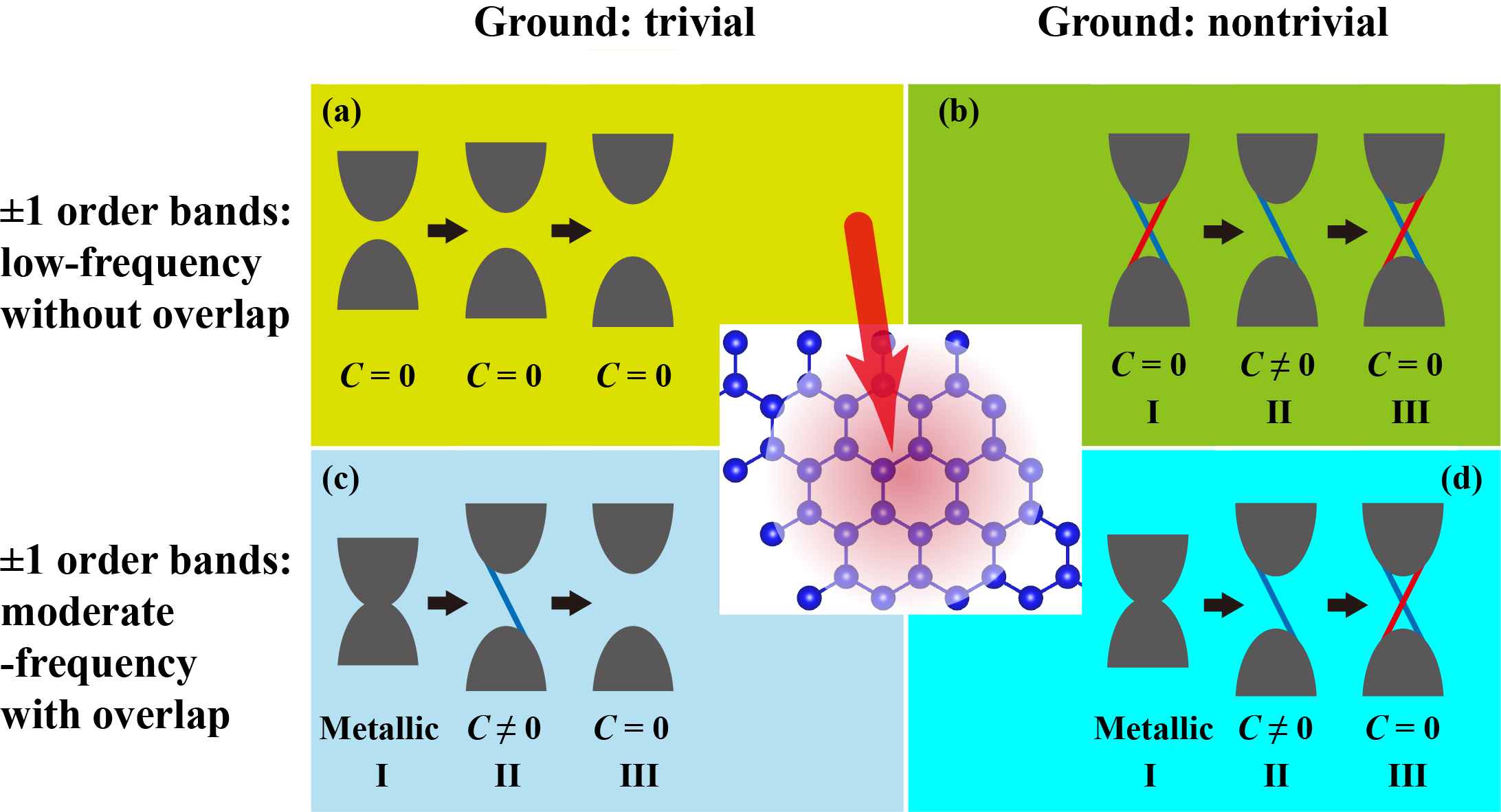}
	\caption{\textbf{Illustration of two-stage TPT process classified by four major conditions of ground state.} \textbf{a} In the top-left section, a initially trivial gapped state maintains its trivial insulating character during the enhancement of light intensity. \textbf{b} The top-right section shows that starting from a nontrivial gapped state (QSH), a QAH state emerges, followed by a compensated Chern insulating state. \textbf{c} In the moderate-frequency regime, starting from a metallic phase, the left-bottom section illustrates a transition from metallic to QAH, then to trivial insulating state, resulting from trivial ground nature. \textbf{d} Conversely, the right-bottom section depicts a transition from metallic to QAH, then to compensated Chern insulating state, arising from nontrivial ground nature. In subfigures \textbf{b}-\textbf{d}, "I", "II" and "III" marked below are corresponding to the three distinct topological phases divided by the two TPT stages. In the center subfigure, a large red arrow indicates the incidence of CPL, with blue spheres representing the Bi or Sb atoms.}
	\label{fig1:Illustration}
\end{figure*}

	At ultralow light intensities (approximately 0.01 Å$^{-1}$), each order of the band structure remains unmodified and splits according to the light frequency. Thus, in the low-frequency regime ( $\hbar\omega$ $\textless$ 0.24 eV), the band structure retains its gapped nature, whereas in the moderate- or high-frequency regime ( $\hbar\omega$ $\geq$ 0.24 eV), a metallic phase emerges at low light intensities. Thereupon, Fig. \ref{fig1:Illustration} illustrates the roadmap of the topological phase diagram unveiled in this work, which is categorized into four distinct circumstances: \textbf{a)} The low-frequency regime begins with a trivial ground state, located in the top-left part of the diagram, resulting in a trivial insulating state. \textbf{b)} The low-frequency regime starts with a QSH ground state, located in the top-right part, transitioning through QSH, QAH, and compensated Chern-insulating states. \textbf{c)} The moderate-frequency regime commences from a metallic and trivial ground state (bottom-left), experiencing transitions through metallic, QAH, and trivial insulating states. \textbf{d)} The moderate-frequency regime starts from metallic but nontrivial ground state (bottom-right), undergoing transitions through metallic, QAH, and compensated Chern-insulating states. Thereby, bismuthene satisfies conditions \textbf{b)} and \textbf{d)}, while for conditions \textbf{a)} and \textbf{c)}, we choose a new material bismuthene-antimonene monolayer, which fulfills the requirement of trivial phase without a significant large gap. In Fig. \ref{fig1:Illustration}, the denoted "I", "II" and "III" stand for the three distinct topological phases: Phase I is for QSH state at $A = 0$ conserved by $T$, and the quasi-QSH state at $A > 0$ without $T$ but before the first gap closure appears; Phase II is pointed to high Chern-number QAH state ($C$ = $\pm$3); Phase III is corresponding to compensated Chern insulating state ($C$ = 0).

	\vspace{1.5em}
	\noindent
	\textbf{Two-stage Phase Transitions}
	
\begin{figure*}
	\centering
	\includegraphics[width=1\linewidth]{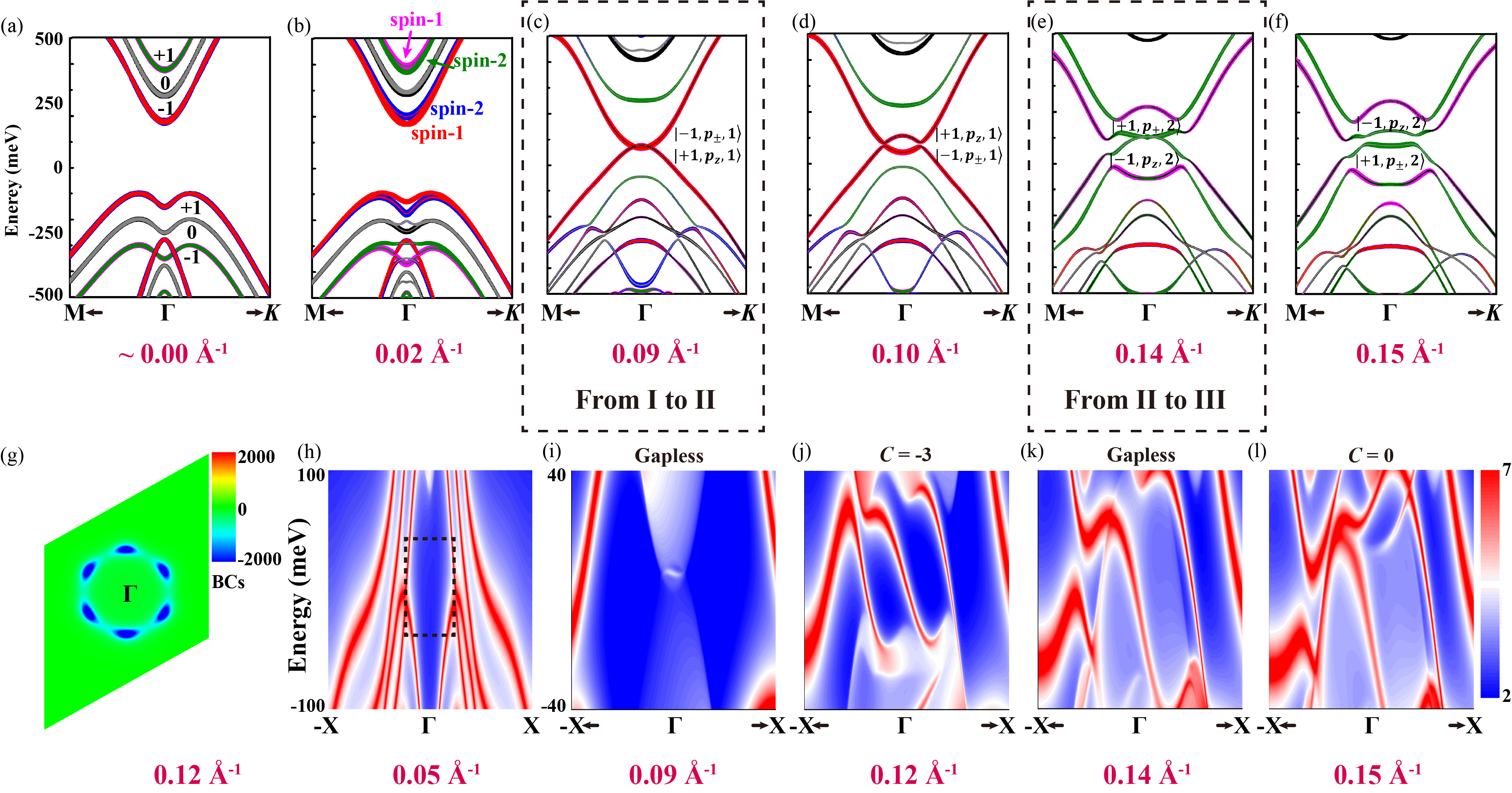}
	\caption{\textbf{Detailed analysis of the two-stage TPT process in R-CPL-irradiated trigonal bismuthene at $\hbar\omega$ = 0.1 eV.} \textbf{a}-\textbf{f} provide a detailed presentation of the two-stage band inversion process. The red and blue bubbles correspond to spin components “\textbf{1}” and “\textbf{2}” respectively, projected by the +1 order replica of the valence band and the -1 order replica of the conduction band. On the contrary, the magenta and olive bubbles are also related to spin “\textbf{1}” and “\textbf{2}” accordingly, but projected by the -1 order replica of the valence band and the +1 order replica of the conduction band conversely. For the zero-order bands, black and gray bubbles reveal the devotions of spin “\textbf{1}” and “\textbf{2}” accordingly. Dashed frames enveloping subfigures \textbf{c} and \textbf{e} highlight the first and the second TPT point. Beneath this, the transitions from Phase I to Phase II and from Phase II to Phase III are also illustrated. \textbf{g} Berry curvature distributions along one BZ at $A$ = 0.12 Å$^{-1}$. \textbf{h} The LDOS pattern across the entire BZ at $A$ = 0.05 Å$^{-1}$. The black rectangular frame in the center indicates the zoomeded-in range for subfigures \textbf{i}-\textbf{l}. The color gradient transitions from blue to white and then to red, representing an increase in the value of LDOS. Subfigures \textbf{i}-\textbf{l} illustrate the evolution of LDOS patterns within the zoomeded-in range at $A$ = 0.09 Å$^{-1}$,  0.12 Å$^{-1}$,  0.14 Å$^{-1}$, and  0.15 Å$^{-1}$, respectively. In subfigure \textbf{a}, "$\sim$ 0.00 Å$^{-1}$" means the light intensity is very weak and near to zero.
	}
	\label{fig2:Two-orderProcess}
\end{figure*}

	\noindent
	For clarity, we mainly focus on the case of  \textbf{b)} in the phase transition diagram. Here, we adopt a low frequency of right-handed CPL (R-CPL, $\hbar\omega$ = 0.1 eV), with the light incident perpendicularly on the bismuthene. Concretely, the band orders are denoted as $ |\left.-1\right\rangle, |\left.0\right\rangle$ and $ |\left.+1\right\rangle$ respectively; the two spin components are represented by $|\left.\textbf{1}\right\rangle$ and $|\left.\textbf{2}\right\rangle$. Figures \ref{fig2:Two-orderProcess}(a)–\ref{fig2:Two-orderProcess}(d) sequentially illustrate the first TPT from Phase I: the QSH state (together with the quasi-QSH state when $T$ is broken), to Phase II: the high Chern-number QAH state. Notably, we focus solely on the eight bands associated with inversion: those colored in red $(|\left.-1,p_\pm,\textbf{1}\right\rangle$, $|\left.+1,p_z,\textbf{1}\right\rangle)$, those in blue $(|\left.-1,p_\pm,\textbf{2}\right\rangle$, $|\left.+1,p_z,\textbf{2}\right\rangle)$,  those in magenta $(|\left.+1,p_\pm,\textbf{1}\right\rangle$, $|\left.-1,p_z,\textbf{1}\right\rangle)$ and those in olive $(|\left.+1,p_\pm,\textbf{2}\right\rangle$, $|\left.-1,p_z,\textbf{2}\right\rangle)$ sequentially. As the light intensity enlarges, the spin degeneracy is lifted, clearly shown in Fig. \ref{fig2:Two-orderProcess}(b) at $A$ = 0.02 Å$^{-1}$. For +1 order of valance band and -1 order of conduction band, the spin "\textbf{1}" (red) gradually shrinks the gap, meanwhile spin "\textbf{2}" (blue) magnifies the gap. Amazingly, the -1 order of valance band and +1 order of conduction band perform oppositely, in which spin "\textbf{1}" (magenta) broadens but spin "\textbf{2}" (olive) reduces the gap. Firstly, The red spin component ($|\left.\textbf{1}\right\rangle$) experiences a band closing-reopening process around light amplitude $A$ = 0.09 Å$^{-1}$ [Fig. \ref{fig2:Two-orderProcess}(c)]. The threefold inversion between the conduction band of -1 order ($|\left.-1,p_\pm,\textbf{1}\right\rangle$) and the valence band of +1 order ($|\left.+1,p_z,\textbf{1}\right\rangle$) governs the first TPT point. Subsequently, in Figs. \ref{fig2:Two-orderProcess}(e) and \ref{fig2:Two-orderProcess}(f), the other TPT process, driven by a threefold band inversion between the conduction band of +1 order ($|\left.+1,p_\pm,\textbf{2}\right\rangle$) and the valence band of -1 order ($|\left.-1,p_z,\textbf{2}\right\rangle$), occurs at a larger light intensity of 0.14 Å$^{-1}$ (dedicated by olive component). Markedly, this band inversion occurs at a different spin component (colored by olive, spin "\textbf{2}"), indicating that the Chern number shifts by the opposite sign compared to the first TPT point. We also paint the zero-order bands with black: $(|\left.0,p_\pm,\textbf{1}\right\rangle)$, and gray: $(|\left.0,p_\pm,\textbf{2}\right\rangle)$, respectively. Remarkably, throughout the entire phase transition process, zero-order band components do not experience any band closure.

    Figure \ref{fig2:Two-orderProcess}(g) illustrates the Berry curvature distributions across one Brillouin zone (BZ), revealing six nonzero regions centered around the $\Gamma$ point. Each nonzero zone contributes a Chern number of --0.5, suggesting the presence of three chiral edge states, with each state connecting two inversion-symmetry-related k points (one corresponding to the conduction band and the other to the valence band). Next, the local density of states (LDOS) patterns projected onto the $Y$ edge are presented as the laser intensity increases. Laser intensities below $A$ = 0.09 Å$^{-1}$ fails to induce any TPT, with 0.05 Å$^{-1}$ selected initially for the complete BZ plot [Fig. \ref{fig2:Two-orderProcess}(h)]. 
    
    To capture the details, we provide zoomed-in images of the LDOS pattern around the $\Gamma$ point, as indicated by the black frame in Fig. \ref{fig2:Two-orderProcess}(h). The first TPT point nearly coincides with 0.09 Å$^{-1}$ [Fig. \ref{fig2:Two-orderProcess}(i)], followed by the emergence of a high Chern-number QAH state ($C$ = -3), with 0.12 Å$^{-1}$ selected as a representative value [Fig. \ref{fig2:Two-orderProcess}(j)]. Further enlarging the light amplitude encounters another band closure and TPT point [Fig. \ref{fig2:Two-orderProcess}(k)], resulting in the emergence of a compensated Chern-insulating phase ($C$ = 0) [Fig. \ref{fig2:Two-orderProcess}(l)]. In addition to the previous nonzero Berry curvature points with negative values, a new sixfold distribution of positive Berry curvature emerges (Supplementary Fig. 3), compensating for the former, ultimately resulting in a zero Chern number state, see particulars in Supplementary Note 3. Supplementary Figure 4 demonstrates the evolution of plane-distributed band gaps, where the TPT point is associated with sixfold band-closing points. It also confirms the presence of threefold band inversions, each occurring at two inversion-symmetry-related points, as described by the trigonal warping model, analogous to that appears at the $K$ valley of monolayer VSi$_2$N$_4$ \cite{zhan2023floquet}.

\begin{figure*}
	\centering
	\includegraphics[width=0.75\linewidth]{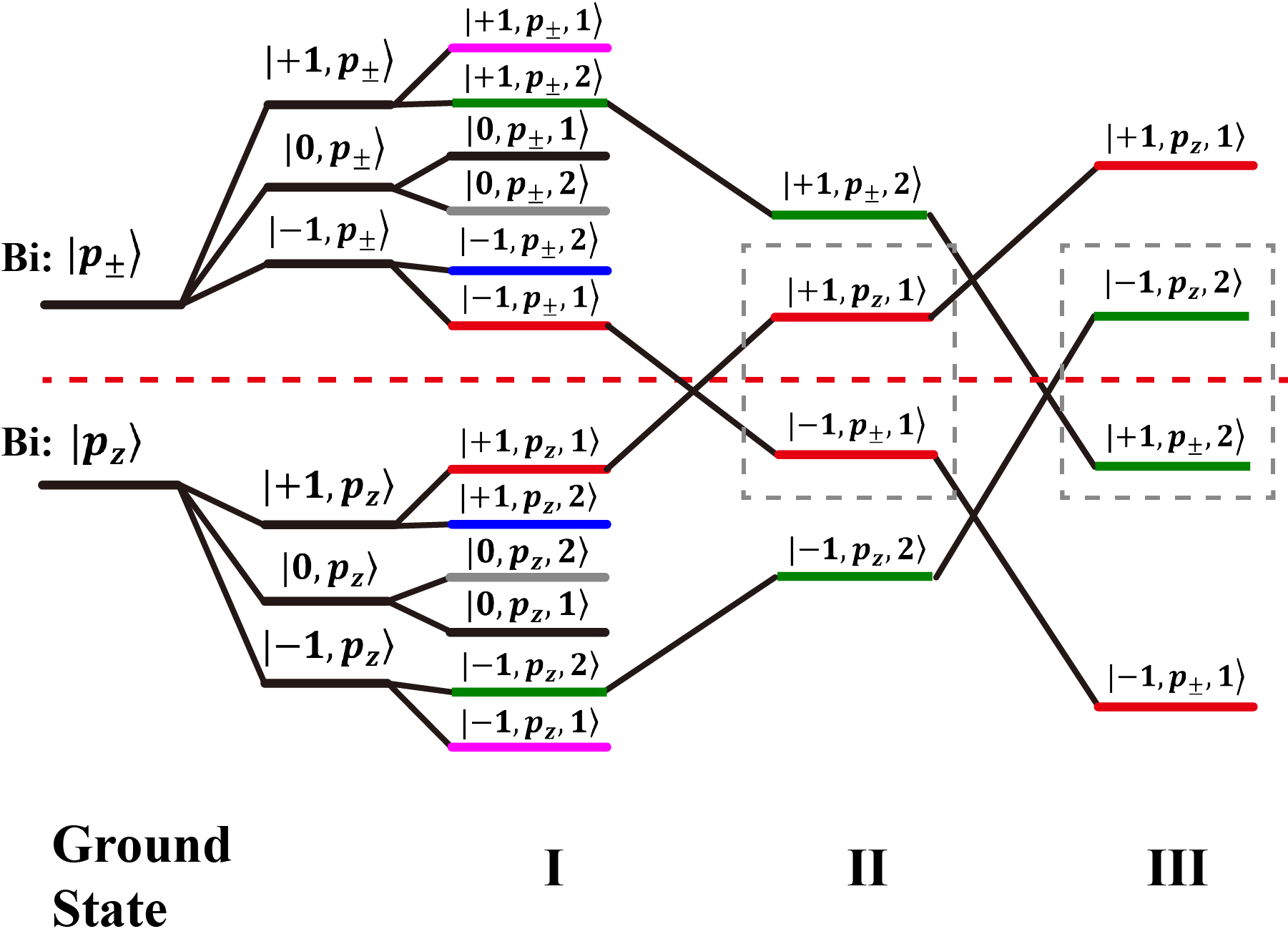}
	\caption{\textbf{Illustration of light-induced topological band-inversion process in trigonal bismuthene-family materials.} The two TPT points divide the entire phase diagram into three distinct phases, designated as Phase I, Phase II, and Phase III. For the +1 order replica of the valence bands and the -1 order replica of the conduction bands, the spin components "\textbf{1}" and "\textbf{2}" are represented by red and blue horizontal lines, respectively. In contrast, for the -1 order replica of the valence bands and the +1 order replica of the conduction bands, the colors are represented by magenta and olive accordingly. For zero-order bands, spin  "\textbf{1}" and "\textbf{2}" are denoted by black and gray horizontal lines. The two gray dashed frames highlight the two band-inversion processes, while the red dashed horizontal line indicates the Fermi energy.}
	\label{fig3:BandInversion}
\end{figure*}

	Accordingly, Fig. \ref{fig3:BandInversion} summarizes the two types of TPTs discussed earlier, along with the band inversion regime. Notably, the zero-order bands are absent from both TPT processes. The spin “$\textbf{1}$” of the -1 order conduction band and the +1 order valence band (colored in red) dominates at the first TPT point. Conversely, the spin “$\textbf{2}$” of the +1 order conduction band and the -1 order valence band (colored olive) governs the second TPT point. Due to the opposing spin components governing these two TPTs, a jump in the Chern number with opposite signs occurs, ultimately resulting in a compensated zero Chern number.

	Analogously discussed in Supplementary Note 4, moderate laser frequencies initially produce a metallic phase; however, a gap subsequently opens, resulting in a Chern-insulating phase (still with $C$ = --3), followed by band closure and the emergence of compensated Chern insulators ($C$ = 0). Similar band topology evolutions are observed in Supplementary Fig. 5 (by selecting $\hbar\omega$ = 0.25 eV), where the trigonal warping term continues to drive the QAH and the compensated Chern-insulating states.

	\vspace{1.5em}
	\noindent
	\textbf{Choosing Laser Parameters: Contour Distributions of Global Gaps and Chern Numbers}
	
\begin{figure*}
	\centering
	\includegraphics[width=0.75\linewidth]{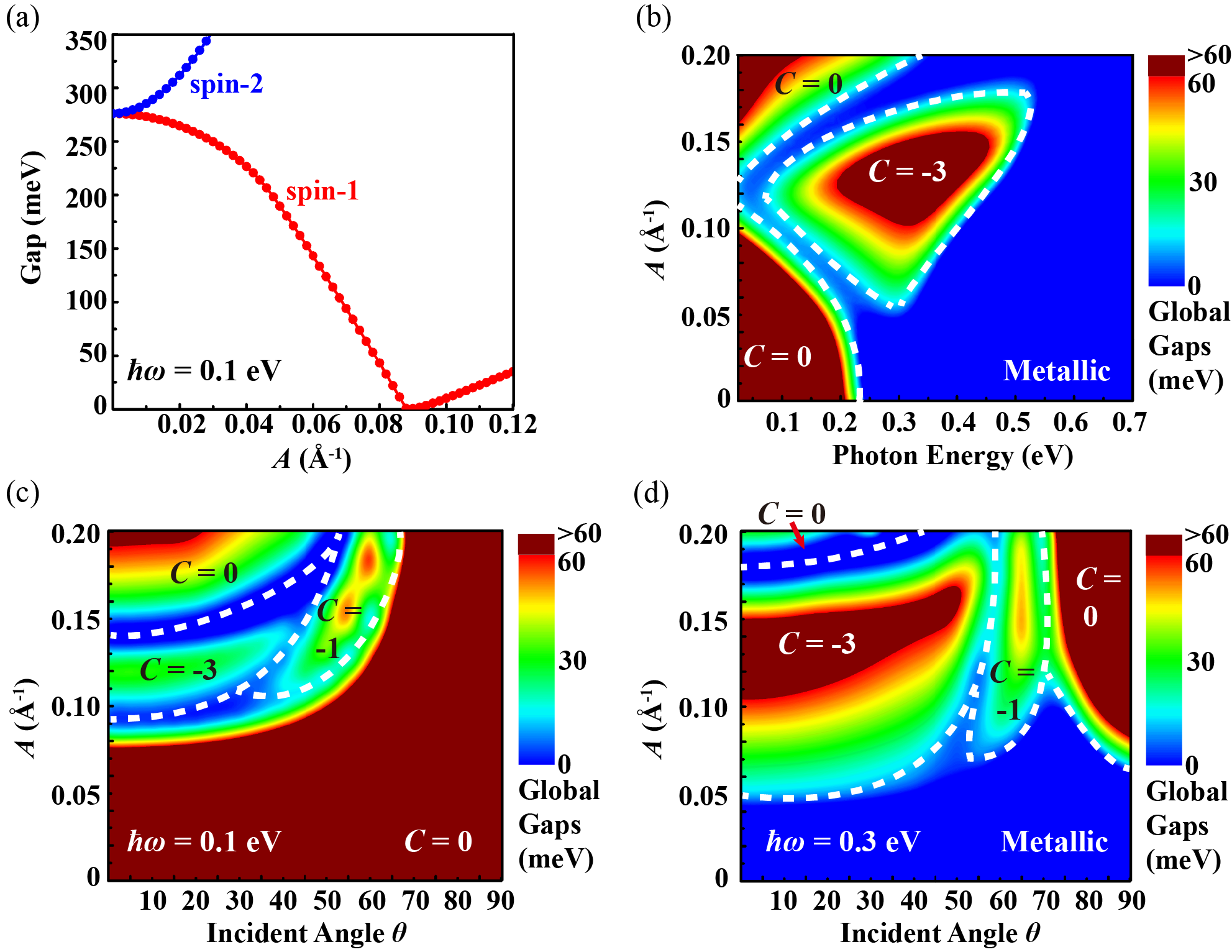}
	\caption{\textbf{Spin splitting and the contour distribution of phase diagrams in R-CPL-induced topologies of trigonal bismuthene.} \textbf{a} Spin-resolved global gap evolution of trigonal bismuthene under the irradiation of R-CPL at  $\hbar\omega$ = 0.1 eV. The red and blue lines correspond to spin "\textbf{1}" and "\textbf{2}", respectively. \textbf{b} Contour distribution of global gaps and Chern number distributions as a function of light frequency ($x$-axis) and intensity ($y$-axis). The color gradient transitions from blue to green and then to red, indicating an enhancement of the global gap. The maroon zone represents the global gap greater than 60 meV, while the metallic zone is denoted by blue. The white dashed curves delineate the boundaries between different Chern number phases. \textbf{c} and \textbf{d} are similar to \textbf{b}, but they illustrate the distributions along the incident angle ($x$-axis) and light intensity ($y$-axis) at $\hbar\omega$ = 0.1 eV and  0.3 eV, respectively.}
	\label{fig4:Contours}
\end{figure*}

	\noindent
	Typically, obtaining a contour distribution diagram provides a key roadmap for selecting laser parameters, including light frequency ($\hbar\omega$), light intensity ($A$), incident angle ($\theta$), and others. Focusing on R-CPL-irradiated bismuthene, spin splitting plays a crucial role in all the Chern-insulating states. Accordingly, in Fig. \ref{fig4:Contours}(a), we compare the gap evolution of the two spin components at $\hbar\omega$ = 0.1 eV. The phase diagram, depicted in Fig. \ref{fig4:Contours}(b), demonstrates the contour distribution of global gaps, accompanied by Chern number distributions delineated by white dashed curves. Noticeably, a triangular-shaped region marked by a non-zero Chern number ($C$ = -3) is located within the laser frequency range of $\hbar\omega$ = 0.05 eV $\sim$ 0.50 eV and the light intensity range of $A$ = 0.05 Å$^{-1}$ $\sim$ 0.18 Å$^{-1}$. In the central region, a maroon area corresponds to ultra-large global gaps ($\geq$ 60 meV), potentially supporting both high Chern-number and beyond-room-temperature QAH states across a wide range of laser parameters. Remarkably, both the low-frequency regime ($\hbar\omega$ = 0.05 eV to 0.24 eV) and the moderate-frequency regime ($\hbar\omega$ = 0.24 eV to 0.50 eV) support QAH states with $C$ = -3. In the high-frequency regime ($\hbar\omega$ $\textgreater$ 0.50 eV), no gap emerges during the whole process up to 0.20 Å$^{-1}$, with the QAH state obscured by the metallic phase (see Supplementary Fig. 6).

	Noteworthily, oblique incident light supplies a greater range of Chern number options, which selectively breaks the $C_{3z}$-rotational symmetry. Figures \ref{fig4:Contours}(c) and \ref{fig4:Contours}(d) respectively display the contour distributions of global gaps and Chern numbers as a function of light intensity ($A$) and incident angle ($\theta$) at $\hbar\omega$ = 0.1 eV and 0.3 eV. As the incident angle exceeds 30$^{\circ}$, a new QAH phase, characterized by $C$ = -1, emerges initially with enhancing light intensity (see Supplementary Fig. 7 of Supplementary Note 5). The light intensity continuing to enhance, the high Chern number of -3 recovers, originating from that one pair of band inversion happens earlier to the other two. This represents a direct prediction of a selectively symmetry-broken based TPT point shifting, which can be manipulated solely by adjusting the orientation of the incident light. Further increasing the oblique angle causes the low-Chern-number zone to fully fill the contour graph up to 0.2 Å$^{-1}$. Beyond 60$^{\circ}$, the QAH state disappears from the graph, leaving only trivial insulating phases. In the moderate-frequency regime ($\hbar\omega$ = 0.3 eV), a similar distribution of two QAH-state regions is observed. Moreover, if the incident light is nearly parallel to the sample, the metallic phase transitions back to a trivial insulating state (see Fig. \ref{fig4:Contours}(d)).
	
	\vspace{1.5em}
	\noindent
	\textbf{Floquet QAH States in Bismuthene Derivatives} 

    \noindent
Despite having the same atomic structure but lower values of SOC strength, bismuthene-antimonene and antimonene monolayers degrade into trivial insulators at the ground state. Nonetheless, with appropriately designed light intensities, Chern insulating features can also be achieved. 

Focusing initially on bismuthene-antimonene, the size of its trivial global gap is moderate ($\sim$ 0.38 eV). A SOC strength of 1.6 times the original value is required to restore its topological character (see Supplementary Fig. 8 in Supplementary Note 6). Figures \ref{fig5:Bismuthene-antimonene}(a) to \ref{fig5:Bismuthene-antimonene}(e) present the corresponding outcomes. Specifically, Fig. \ref{fig5:Bismuthene-antimonene}(a) shows a top view of bismuthene-antimonene, while Figs. \ref{fig5:Bismuthene-antimonene}(b) to \ref{fig5:Bismuthene-antimonene}(d) illustrate the two TPT processes: from metallic to QAH ($C$ = -3), and then to trivial insulating states, by employing light frequency as $\hbar\omega$ = 0.3 eV with zoomed-in LDOS patterns. Subsequently, a contour distribution diagram is presented, confirming the presence of a small, horsebean-like region that also supports the QAH state as described by the trigonal warping model [Fig. \ref{fig5:Bismuthene-antimonene}(e)]. The mean value of global gap reaches about 20 meV, while still maintaining high-temperature QAH states. 

Disappointingly, monolayer antimonene fails to establish a Chern-insulating state at any position on the contour-distributed phase diagram [Fig. \ref{fig5:Bismuthene-antimonene}(j)], transitioning directly from a metallic phase to a trivial insulating character in the high-frequency regime ($\hbar\omega$ $\textgreater$ 0.5 eV). Despite the surprising finding that the light-induced QAH state does not necessarily arise from a topologically nontrivial state, it is crucial to ensure that the initial band gap avoids excessively large values. Hence, compared to one that originates from a topologically nontrivial state, light-induced QAH state from a trivial insulator imposes significantly more stringent requirements on laser parameters.

Other additional information is available in the Supplementary Materials, with the main conclusions summarized below. Seen from Supplementary Note 7, reversing the chirality of CPL (i. e., left-handed CPL) yields similarly two-stage TPT process, but with an opposite sign for the QAH state ($C$ = +3), as shown in Supplementary Fig. 9. Unfortunately, linearly polarized light cannot induce Chern insulating states; however, it can introduce a global gap from the metallic state (see Supplementary Fig. 10). For bismuthene, gradually and artificially reducing the SOC strength (see Supplementary Note 8) drives its ground state back to a trivial phase ($\leq$ 40 \%), but the region of the QAH state remains robust (see Supplementary Figs. 11-12). As such, varying the SOC strength of bismuthene-antimonene also preserves the high Chern-number QAH-state island in the phase diagrams (Supplementary Fig. 13). Besides, bismuthene grown on suitable substrates, such as H-saturated SrS and SrSe, maintains the aforementioned properties (see Supplementary Figs. 14-15), with detailed analysis exhibited in Supplementary Note 9.
	
	\vspace{1.5em} 
	\noindent
	\textbf{\large Discussion}
	
	\noindent
	For the laser parameters that uphold high Chern-number QAH states, the maximum light intensity and frequency reach approximately $A$ = 0.18 Å$^{-1}$ and $\hbar\omega$ = 0.5 eV correspondingly. Based on insights from previous experimental investigations that indicate the creation of $\pm$1 order replica bands occur after six cycles of incident light \cite{qiu2018ultrafast}, the maximum fluence reaches approximately 0.005 J/cm$^{2}$. Experimental analyses of damage in bismuth indicate a threshold of 0.022 J/cm$^{2}$ \cite{misochko2004observation}, significantly higher than the value utilized in this study. Several related investigations also support that thermal damage can be completely avoided under our laser parameters. For example, monolayer square bismuth has been computationally tested up to 0.5 Å$^{-1}$ \cite{zhang2025quantum}, much greater than the maximum light intensity employed in this work. Similar light-intensity selections (0.5 Å$^{-1}$) have also been implemented in theoretical predictions for tensile-strained monolayer Bi(110) and Bi(111) \cite{kong2024tunable}. All these assessments affirm the safety of the laser parameters employed in this study.

	In conclusion, we firstly present a guiding roadmap for light-induced TPTs in $\Gamma$-point-based topology, using trigonal bismuthene and its derivatives as templates. Irradiated by low or moderate frequencies of CPL, a two-stage TPT process occurs: first, a QSH or metallic phase emerges, followed by the formation of a high Chern-number QAH state ($C$ = -3), and ultimately a compensated Chern insulating phase ($C$ = 0) appears. Both Chern-insulating phases are described by a threefold band inversion within the trigonal warping model. The $\pm$1 order replicas of the top valence and bottom conduction bands contribute to the aforementioned TPTs. The high Chern-number QAH state remains robust across a wide range of laser parameters, exhibiting substantial global gaps that support applications beyond room-temperature. Furthermore, QAH state is not indispensably derived from QSH state, but can also be induced from small gapped trivial phase. Our findings offer critical insights into manipulating traditional $\Gamma$-point-based topology through Floquet engineering, highlighting potential future applications in electronic devices.

\vspace{1.5em} 
\noindent
\textbf{\large Methods}

\noindent
\textbf{First-principles Calculations of Ground State Electronic Structure}

\noindent
The ground states of bismuthene, bismuthene-antimonene, antimonene and their heterostructures grown on substrates were obtained using first-principles computational methods implemented in the Vienna Ab-initio Simulation Package (VASP) \cite{kresse1996efficient}, supported by constructing a tight-binding Hamiltonian (TBH). For the k-point mesh grids, a 11$\times$11$\times$1 grid was adopted for the relaxation and self-consistent calculations for all above three monolayers. Among aforementioned building blocks, we applied the criterion of structural optimization, ensuring that the Hellmann-Feynman force on each atom was smaller than 0.001 eV/Å. For the electron energy convergence criterion, a threshold of 1.0$\times$10$^{-7}$ eV was selected for relaxations, self-consistent calculations, and band structure computations. We implemented the PBE functional to extract their ground states \cite{perdew1996generalized}. During the structural relaxation step, SOC effects were not considered; however, SOC effects were included in the self-consistent and band structure computation steps. Additionally, a vacuum layer of at least 15 Å was included to mimic two-dimensional behavior \cite{tkatchenko2009accurate} among the related building-blocks. Given the monolayer structural nature, the van der Waals corrections \cite{grimme2010consistent} were only employed when the bismuthene is put upon substrates. Finally, the VASPKIT \cite{wang2021vaspkit} and PHONOPY \cite{togo2015first} codes were utilized for the post-processing of the computed data.

\vspace{1.5em}
\noindent
\textbf{Simulating Irradiation of Periodic Laser Fields}

\noindent
To obtain the excited states under periodic optical fields, we first employed the Wannier90 package to construct the TBH of all the related materials based on maximally localized Wannier functions \cite{mostofi2014updated,marzari1997maximally,souza2001maximally}. After acquiring the ground-state TBHs, we employed our group-developed code to apply the Peierls substitution \cite{oka2019floquet}: 

\begin{equation}
	H_{ij}(t) = H_{ij} \exp\left[ -\mathrm{i} \int_{R_j}^{R_i} \textbf{\textit{A}}(t) \cdot d\mathbf{r} \right]
	\label{eq1}
\end{equation}

In eq. \ref{eq1}, $R_i$ and $R_j$ stand for the two sites of Wannier charge center.

Concretely, the displacement vector can be defined as:

\begin{equation}
	\begin{split}
		d_{ij} = L_k+w_i-w_j
	\end{split}
	\label{eq2}
\end{equation}

In eq. \ref{eq2}, $w_{i}$ and $w_{j}$ are corresponding to the two sites of Wannier charge center. $L_k$ is the Cartesian coordinate of the $k$-th lattice point.  

The amplitude of the optical vector potential $A_{ij}$ can be expressed as:

\begin{equation}
	\begin{split}
		A_{ij} & = \Biggl[  \left( A_x d_{ij,x} \sin \phi_1 + A_y d_{ij,y} \sin \phi_2 + A_z d_{ij,z} \sin \phi_3 \right)^2 \\
		& + \left( A_x d_{ij,x} \cos \phi_1 + A_y d_{ij,y} \cos \phi_2 + A_z d_{ij,z} \cos \phi_3 \right)^2 \Biggr]^{1/2}
	\end{split}
\end{equation}

$A_x$, $A_y$ and $A_z$ are light intensities along three axes, while $\phi_1$, $\phi_2$ and $\phi_3$ are phase parameters of the light.

The phase angle $\phi_{ij}$ forms as:

\begin{equation}
	\phi_{ij}  = 
	\tan^{-1}\left( \frac{A_x d_{ij,x} \sin \phi_1 + A_y d_{ij,y} \sin \phi_2 + A_z d_{ij,z} \sin \phi_3}{A_x d_{ij,x} \cos \phi_1 + A_y d_{ij,y} \cos \phi_2 + A_z d_{ij,z} \cos \phi_3}  \right)
\end{equation}

After the excitation of the light, the expanded Hamiltonian can be expressed as:

\begin{equation}
	H_{ij}^{(q)} = H_{ij}^{(k)} \textit{e}^{\mathrm {i} q \phi_{ij}} J_q \left( \frac{e}{\hbar} A_{ij} \right)
\end{equation}

$H_{ij}^{(k)}$  represents the original Hamiltonian derived from the Wannier functions, while $J_q$
denotes the $q$-th order Bessel function that characterizes the tunneling effect induced by photons. The variable $q$ indicates the number of photons that are either emitted or absorbed. In this study, $q$ takes on the values -2, -1, 0, +1, and +2. The $\pm2$ order replica bands are not present in the final band structures; instead, they are included solely to introduce perturbations to the zero-order bands and the $\pm1$ order replica bands.

After obtaining the Floquet-engineered TBH, we employed the Green's function method using the WannierTools package \cite{wu2018wanniertools} to investigate the topological features of edge states, Berry curvature distributions, and Chern numbers.

	\vspace{1.5em} 
	\noindent
	\textbf{\large Data Availability}
	
	\vspace{0.5em} 
	\noindent
	The data that support the findings of this work are available from the manuscript's main text and Supplementary Materials. The additional data is available from the corresponding author upon reasonable request.

\vspace{1.5em} 
\noindent
\textbf{\large Code Availability}

\vspace{0.5em} 
	\noindent
	The self-developed codes that implement the optical field are available from the corresponding author upon reasonable request. The ground state computations have been performed by VASP, Wannier90, WannierTools, PHONOPY and VASPKIT codes which can be requested from the corresponding developers.

\vspace{1.5em} 
\noindent
\textbf{\large Acknowledgements}

\vspace{0.5em} 
	\noindent
	We thank Prof. Huisheng Zhang, Dr. Feng Xue, Dr. Hui Zhou, Xiyu Hong, Jingjing Cao for helpful discussions, and thank Dr. Huixia Fu, Dr. Hang Liu for technical supports. The numerical calculations have been done on the supercomputing system in the Huairou Materials Genome Platform.

\vspace{1.5em} 
\noindent
\textbf{\large Funding}

\vspace{0.5em} 
	\noindent
	This work is supported by National Natural Science Foundation of China (No. 12025407 and No. 12450401), National Key Research and Development Program of China (No. 2021YFA1400201), and Chinese Academy of Sciences (No. YSBR-047 and No. XDB33030100). 

\vspace{1.5em} 
\noindent
\textbf{\large Author Contributions}

\vspace{0.5em} 
	\noindent
	Z. L. undertook the majority works of computations and data-processing. H. C. and Z. L. conducted the fundamental theoretical analysis and discussions. S. M. guided the whole project and participated in the design and data analysis of the project. The paper was written by Z. L., H. C. and S. M.  All authors contributed to the discussions and analyses of the data, and approved the final version.

\vspace{1.5em} 
\noindent
\textbf{\large Competing Interests}

\vspace{0.5em} 
	\noindent
	The authors declare no competing interests.

\vspace{1.5em} 
\noindent
\textbf{\large References}
\bibliography{Main_text}
	
\begin{figure*}
	\centering
	\includegraphics[width=1\linewidth]{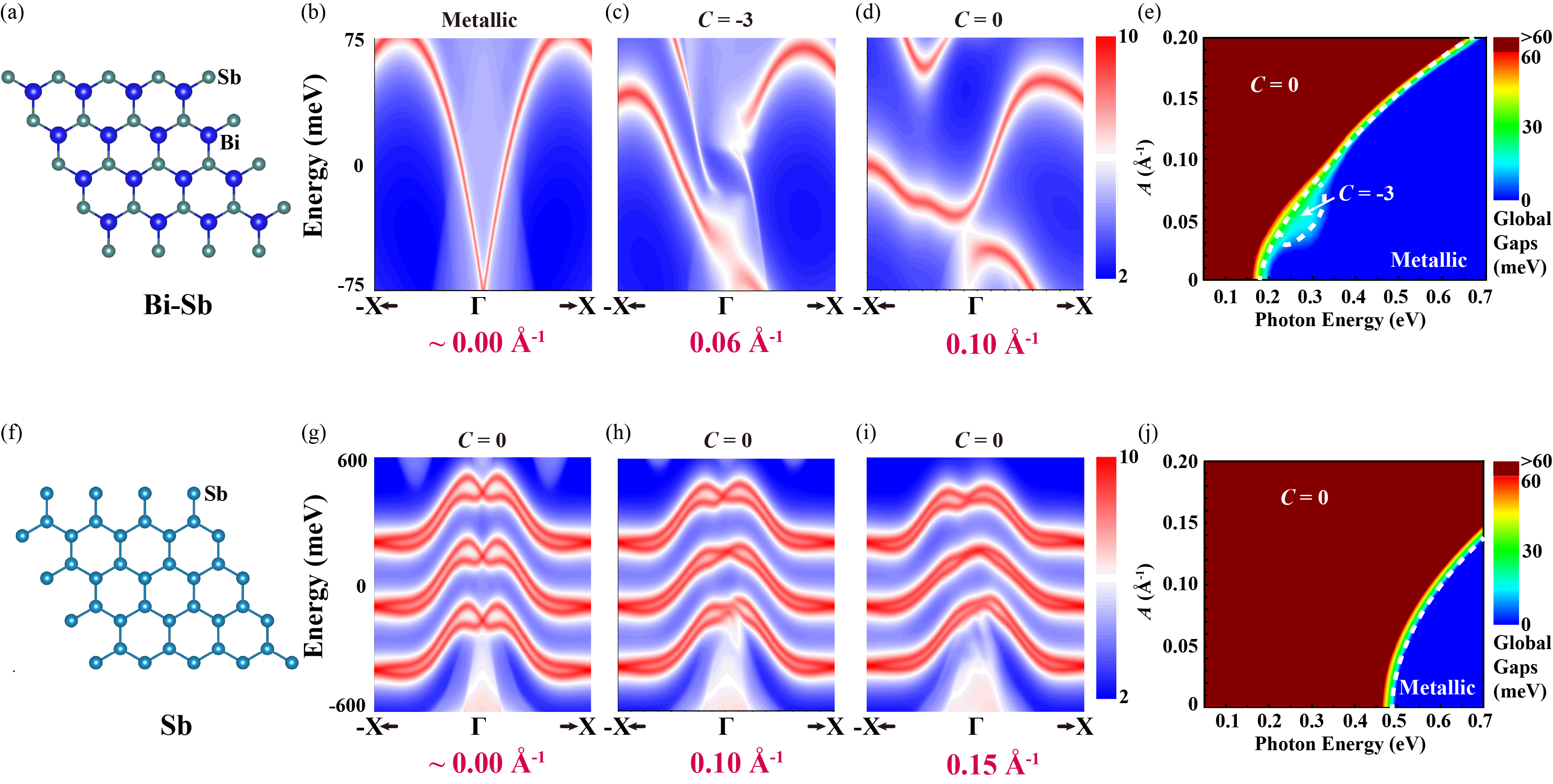}
	\caption{\textbf{Structures and TPTs based on R-CPL-irradiated trigonal bismuthene-antimonene and antimonene.} \textbf{a} The top view structure of trigonal bismuthene-antimonene is shown, with dark-green and blue spheres representing Sb and Bi atoms, respectively. \textbf{b}-\textbf{d} present zoomed-in LDOS patterns of bismuthene-antimonene under the irradiation of R-CPL at $\hbar\omega$ = 0.3 eV. The selected light intensities are $\sim$ 0.00 Å$^{-1}$, 0.06 Å$^{-1}$, and 0.10 Å$^{-1}$, respectively. \textbf{e} displays the contour distribution of global gaps and Chern numbers as a function of laser frequency ($x$-axis) and light intensity ($y$-axis). \textbf{f}-\textbf{j} are analogous to \textbf{a}-\textbf{e}, but focus on the structure of trigonal antimonene. In subfigures \textbf{g}-\textbf{i}, the selected light intensities are $\sim$ 0.00 Å$^{-1}$, 0.10 Å$^{-1}$, and 0.15 Å$^{-1}$ sequentially. "$\sim$ 0.00 Å$^{-1}$" means the light intensity is very weak and near to zero.}
	\label{fig5:Bismuthene-antimonene}
\end{figure*}

\end{document}